\documentclass[showpacs,preprintnumbers,amsmath,amssymb,prl,twocolumn]{revtex4}

\usepackage{graphicx}
\usepackage{dcolumn}
\usepackage{bm}
\def\comment#1{}

\begin{document}

\title{Deconfinement
transition in three-dimensional  compact $U(1)$ gauge theories coupled
to matter fields
}
\author{Hagen Kleinert}
\email{kleinert@physik.fu-berlin.de}
\homepage{http://www.physik.fu-berlin.de/~kleinert/}
\author{Flavio S. Nogueira}
\email{nogueira@physik.fu-berlin.de}
\affiliation{Institut f\"ur Theoretische Physik,
Freie Universit\"at Berlin, Arnimallee 14, D-14195 Berlin, Germany}
\author{Asle Sudb{\o}}
\email{asle.sudbo@phys.ntnu.no}
\affiliation{Department of Physics, Norwegian University of
Science and Technology, N-7491 Trondheim, Norway}

\date{Received \today}

\begin{abstract}
It is shown that permanent confinement
in three-dimensional compact U(1) gauge theory 
can be destroyed by matter fields in
a deconfinement transition.
This is a consequence of a non-trivial infrared fixed 
point caused
by matter, and
an anomalous scaling dimension of the gauge field.
{This leads to
a logarithmic interaction between the defects  of the gauge-fields,
which form a gas of
{\em magnetic\/} monopoles}.
 In the presence of  logarithmic   interactions,
 the original
 {\em electric\/} charges are {\em unconfined\/}.
The
 confined phase,
which is permanent in the absence of matter fields,
is reached at a
   critical electric charge,
where the 
interaction
between magnetic
charges
is screened
by a pair unbinding transition
in a
Kosterlitz-Thouless type of phase-transition.
 \end{abstract}

\maketitle

In a seminal paper, Polyakov
\cite{Polyakov} has shown that
QED in $d=3$ dimensions  with 
a compact $U(1)$ gauge field
exhibits confinement
of electric charges for all values of the coupling constant.
 The origin of this behavior lies in the fact that  the
defects of the gauge field
defined by the boundaries
of surfaces where the gauge field $A_\mu$ jumps by $2\pi$
form a gas of
magnetic monopoles, whose initially long-range
interaction
is reduced to a short range interaction
by Debye  screening.
This screening gives the
initially unobservable
jumping surfaces an energy
leading to an area law for the Wilson integral
and thus to permanent confinement between electric charges \cite{KBook}.

An important question is whether this behavior is changed
by matter fields. The answer is particularly relevant
for  present-day condensed-matter physics,
 where
the effective actions assumed to govern
 strongly correlated electrons
contain a compact $U(1)$ gauge field coupled to matter
\cite{Nayak,Nagaosa,Kim}.
This makes
 the confinement properties
of three-dimensional euclidean gauge-theories
 relevant
for the quantum properties of
strongly correlated
electrons
at zero temperature in {\it two} spatial dimensions.
The existence of a
 confinement-deconfinement (CD) transition in
 gauge-theories with matter has been suggested
to offer an explanation for
a spin-charge separation transition
of slave particles
in the electron system
\cite{Mudry,Laughlin,Nayak,Kim,Senthil,Nagaosa}.

It has
  been  argued
that the presence of matter fields
should {\em not\/}
destroy the
permanent confinement in compact $U(1)$ gauge
theories when the matter field carries a fundamental charge
\cite{FradShe,Nagaosa}, but there is no
universal agreement on this point \cite{Savit,Matsui}. In
this Letter we shall argue that the coupling to such matter fields
induces an anomalous scaling dimension to the gauge field, 
which indeed may give rise to a CD transition in three dimensions.

We discuss first the case of bosonic matter
in a Ginzburg-Landau (GL) model of
superconductivity (denoted Higgs model in particle physics).
We show how a
CD transition
arises at a certain Ginzburg parameter $\kappa$, which is the ratio between
magnetic penetration depth  and coherence length.

In the case of
fermion matter, we consider QED with $N$ four-component Dirac fermions.
Such a system is believed to describe the low-energy
behavior of a  quantum Heisenberg
antiferromagnet (QHA) around the mean-field flux phase
\cite{Marston,Affleck,Kim}. For this model the situation is less clear 
due to spontaneous chiral symmetry breaking.  
In principle  electric charges
are permanently confined below a critical value of $N$.

For a noncompact gauge field,
the   GL Lagrangian reads

\begin{equation}
\label{bosons}
{\cal L}_b=\frac{1}{4e_0^2}F_{\mu\nu}^2+
|(\partial_\mu+iA_\mu)\phi|^2+V(|\phi|^2),
\end{equation}
where  $F_{\mu\nu}=\partial_\mu A_\nu-\partial_\nu A_\nu$ and
$V(|\phi|^2)=-m_0^2|\phi|^2+u_0|\phi|^4/2$.
For the sake of the  discussion
to follow we
assume the field $\phi$ to have $N/2$ complex components
such that the theory is O($N$)-symmetric.
The superconductor has $N=2$.
The
critical behavior of this model
 has been extensively discussed in the
literature.
 Let us summarize the most important properties
for the present discussion.
Traditional
RG calculations such as $\epsilon$-expansion fail to show
a non-trivial fixed point for $N=2$ \cite{HLM}. This
result seems to be an artifact of the $\epsilon$-expansion,
since
a non-trivial fixed point has been demonstrated to exist
\cite{Herbut,Hove,Various,tri,KleinNog}.
The Ginzburg parameter
$ \kappa $ where a fixed point appears first
has been located by a duality transformation
in Ref.~\cite{tri}
and confirmed in  recent large-scale Monte Carlo 
simulations \cite{Mo}. In the GL model itself,
a non-trivial
infrared stable fixed point has  been
found recently by working
in $d=3$ dimensions in the {\em ordered\/} phase \cite{KleinNog}.

We introduce the dimensionless renormalized couplings
$\alpha=e^2\mu^{d-4}$ and $g=u\mu^{d-4}$, where $\mu$ is the
running mass scale. Gauge invariance implies that
$e^2=Z_A e_0^2$ \cite{ZJ}, where $Z_A$ is the gauge field wave 
function renormalization. Thus, we obtain the $\beta$-function 
for $\alpha$:

\begin{equation}
\label{betaalpha}
\beta_\alpha(\alpha,g)=[\gamma_A(\alpha,g)+d-4]\alpha,
\end{equation}
where we have defined the RG function
$\gamma_A\equiv\mu\partial\ln Z_A/\partial\mu$. If a non-trivial
infrared stable
fixed point exists, it must satisfy the equations
$\gamma_A(\alpha_*,g_*)=4-d$ and $\beta_g(\alpha_*,g_*)=0$,
where $\beta_g$ is the $\beta$-function for the $g$-coupling. The
anomalous dimension of the gauge field is such that the
critical correlation function
$\langle A_\mu(p)A_\nu(-p)\rangle=D(p)(\delta_{\mu\nu}-p_\mu p_\nu/p^2)$
with $D(p)\sim 1/p^{2-\eta_A}$ at
large distances. The existence of the infrared stable fixed
point implies that $\eta_A$ is determined exactly:
$\eta_A=\gamma_A(\alpha_*,g_*)=4-d$, for all dimensions
$d\in(2,4)$ \cite{Herbut}.
This exact result was confirmed by Monte Carlo simulations
in conjunction with duality arguments \cite{Hove},
providing further evidence for the existence of the non-trivial
fixed point.
The exact result $\eta_A=4-d$ is of great importance
for the scaling behavior of physical quantities in the
superconductor \cite{Herbut,Nogueira}.
We now consider  how it
affects the physics of the CD transition.
 
The result $\eta_A=4-d$ implies that $D(p)\sim 1/|p|^{d-2}$, or
$D(x)\sim 1/|x|^2$ in real space, for all $d\in(2,4)$. Hence,
the scaling dimension of the gauge field is one for
this range of dimensionality. Consequently, the Maxwell term
is irrelevant in the RG sense if $d\in(2,4)$.
We
emphasize that this corresponds to an {\em exact\/}
 behavior of the theory which
is {\em independent\/} of perturbation theory.
The result amounts to an
 effective Lagrangian for the gauge field

\begin{equation}
\label{effec}
{\cal L}_A=\frac{1}{4\alpha_*}
F_{\mu\nu}\frac{1}{(-\partial^2)^{\eta_A/2}}
F_{\mu\nu}.
\end{equation}

The anomalous scaling leads to a
potential between two test charges at equal times
given by

\begin{equation}
\label{potential}
V(R)\sim \frac{1}{R^{d-3+\eta_A}}\sim\frac{1}{R},
\end{equation}
for all $d\in(2,4)$.
If the anomalous
dimension were zero, we would obtain a
 behavior
$V(R)\sim\ln R$ in $d=3$.
The behavior (\ref{potential})
corresponds to an unconfined
 Coulomb gas in the theory.
There exists a similar phase
in supersymmetric QCD in four dimensions, where
the $\beta$-function is known exactly \cite{Novikov}. It has been
argued by Seiberg \cite{Seiberg} that this theory has a
non-trivial infrared stable fixed point for all
$N_f\in(3N_c/2,3N_c)$, where $N_f$ and $N_c$ are the number of
flavors and colors, respectively.
In this range of $N_f$, the quarks and
gluons are interacting {\it massless} particles which are
not confined.

We next account for the compact nature  of the gauge field
which gives rise
to
magnetic monopoles producing
confinement.
In general,
a coupling to
matter fields weakens confinement.
The competition between the two in principle could
lead to a CD transition.
The compact nature
of the gauge fields is
most easily accounted for
by
the introduction of
so-called {\em plastic gauge fields\/} \cite{KBook,Cam}
$n_{\mu\nu}$ which are superpositions of
$ \delta $-functions on surfaces over which the
angular vector field components $A_\mu$  jump by $2\pi$:

\begin{equation}
\label{effecmon}
{\cal L}_{A}'=\frac{1}{4\alpha_*}
(F_{\mu\nu}-2\pi n_{\mu\nu})\frac{1}{\sqrt{-\partial^2}}
(F_{\mu\nu}-2\pi n_{\mu\nu}).
\end{equation}
Here, we have specialized to $d=3$ and thus $\eta_A=1$.
From the $n_{\nu \mu}$, we obtain the monopole density:
\begin{equation}
\label{monopoles}
\epsilon_{\mu\nu\lambda}\partial_\mu n_{\nu\lambda}\equiv m(x)=\sum_a q_a\delta^3(x-x_a),
\end{equation}
where $q_a=\pm$integer are the monopole charges.
By a duality transformation,
the partition function
associated with
(\ref{effecmon}) can be brought to the
equivalent form

\begin{widetext}
\begin{equation}
\label{effecmonx}
Z=\sum_{\{m(x)\}}\int{\cal D}\chi
 \exp\left\{\int d^3x\left[
-\frac{ \alpha _*}{2}
( \partial_\mu \chi  )\sqrt{-\partial^2}
 (\partial_\mu  \chi)
-2\pi i m(x)\chi(x)\right]\right\},
\end{equation}
\end{widetext}
where $\chi(x)$ is the dual electromagnetic potential
which is a scalar in three dimensions.
It can be integrated out to yield a monopole gas
with a partition function
\begin{equation}
Z=\sum_{{\rm mon. configs.}}\exp\left[ -\frac{2\pi^2}{\alpha_*}
\sum_{a,b}q_a q_bV(x_a-x_b)\right] ,
\label{@}\end{equation}
with the potential
 $V(x)=\int d^3k\,e^{ikx}
/(2\pi)^3|k|^3
$.
This is a logarithmic potential
in three dimensions.
The monopoles have a
large self-energy and thus a low fugacity
 $ \zeta $. We may therefore  restrict the sum  to
 $q_a=\pm 1$,
where
 (\ref{effecmonx}) reduces to
the following sine-Gordon-like partition function in three dimensions
\begin{equation}
\label{SG}
Z\approx
\int {\cal D}\chi
e^{ -\frac{1}{2t}
\int
 d^3x\left[\chi(-\partial^2)^{3/2}\chi-z\cos\chi\right]} ,
\end{equation}
where $z=8\pi^2\zeta/\alpha_*$,
$t=4\pi^2/\alpha_*$. The above treatment closely 
parallells that of Polyakov for
pure compact QED \cite{Polyakov}, the novel 
result being the appearance of the
anomalous gradient term due to the presence of 
matter fields.
This anomalous gradient term is in contrast to the usual
U(1) gauge theory
where it has the standard
 $\chi(-\partial^2)\chi$ and receives
a mass from the
$\cos \chi $ term causing permanent  confinement
of electric charges.

Remarkably, the logarithmic behavior caused by
the
anomalous gradient term
gives 
 rise to  a CD phase transition in three dimensions
driven by a magnetic monopole-antimonopole
unbinding transition,
very
similar to a Kosterlitz-Thouless vortex-antivortex 
unbinding 
transition in two dimensions.
Its position is governed by the precise value of 
$ \alpha _*$
which depends on
 $N$.
By calculating the classical expectation
value of the dipole moment of a single pair
$\langle r^2\rangle $ it is easy to see
 that the KT-like
pair separation  transition
occurs at $t=t_c=12\pi^2$. For
$t<t_c$, the field
$\chi$ is massive and electric charges are  
confined.

In the ordered phase the system has two length scales whose
ratio gives the Ginzburg parameter $\kappa$, which in turn can
be written as $\kappa=\sqrt{g/2\alpha}$. Thus, we see that the
theory can be parametrized in terms of $\alpha$ and $\kappa$,
instead of $\alpha$ and $g$. In such a situation it is more
convenient to use the Higgs mass as the running scale, i.e. 
$\mu=m$.
At one-loop level, the RG function $\gamma_A$ in the ordered phase 
for $d=3$ is given by  \cite{KleinNog}

\begin{equation}
\label{gammaA}
\gamma_A=\frac{\sqrt{2}C(\kappa)\alpha}{24\pi(2\kappa^2-1)^3},
\end{equation}
where $C(\kappa)=4\kappa^6+10\kappa^4-24\sqrt{2}\kappa^3
+27\kappa^2+4\sqrt{2}\kappa-{1}/{2}$. The non-polynomial form
in $\kappa$ of $\gamma_A$ comes from the fact that $\kappa=m/m_A$,
where $m_A$ is the gauge field mass. 
Details of the derivation can be found in Ref. \cite{KleinNog}.

An effective gauge coupling
$\bar{\alpha}(\kappa)$
can be defined by the solution of the equation
$\gamma_a(\bar{\alpha},\kappa)=1$, which gives a
critical line. This critical line makes sense only for
$\kappa>1/\sqrt{2}$, that is, in the type II regime, or in
the interval $0\leq  \kappa<0.096/\sqrt{2}$ \cite{KleinNog}
deep in the type I regime. In the interval
$0.096/\sqrt{2}<\kappa<1/\sqrt{2}$ the RG function $\gamma_A$ is
negative, which means that {\it the theory is asymptotically free
in this interval}. This is a remarkable result for an Abelian
theory. It cannot be obtained with standard perturbation theory
using the $\epsilon$-expansion, but is easily obtained by
performing a one-loop calculation in the ordered phase and $d=3$. Typically,
ordinary perturbation theory can access only the deep type I regime.
Note that $\bar{\alpha}(\kappa)\to\infty$ as
$\kappa\to 0.096/\sqrt{2}$ from the left. This means that near
$\kappa=0.096/\sqrt{2}$ perturbation theory breaks down. Remarkably,
perturbation theory can be trusted in the type II regime sufficiently
close to $\kappa=1/\sqrt{2}$ where $\bar{\alpha}$ is
small \cite{KleinNog}.
We stress that all
these results are made possible only because 
there exist two mass scales in the ordered phase. 

If we now use the critical coupling
$t_c$ of our sine-Gordon-like
 theory
$4\pi^2/\bar{\alpha}(\kappa)$,
we find that the
Kosterlitz-Thouless-like phase transition takes place
at $\kappa_c=1.17/\sqrt{2}$,
{\it which is precisely the fixed-point value of }
$\kappa$ {\it obtained from the zero of the} $\beta$-{\it function }
$\beta_\kappa\equiv m\partial\kappa/\partial m$
{\it in the non-compact theory} \cite{KleinNog}. 
This result is consistent with the scenario that there is no phase
boundary between the Higgs and the confining phase when the matter
field carries the fundamental charge  \cite{FradShe,OstSeil}.

The coupling to fermionic matter fields will now be considered.
The Lagrangian is given by

\begin{equation}
\label{Dirac}
{\cal L}_f=\frac{1}{4e_0^2}F_{\mu\nu}^2+
 \sum_{i=1}^N\bar{\psi}_i\gamma_\mu(\partial_\mu+iA_\mu)\psi_i.
\end{equation}
The above Lagrangian corresponds to an effective theory for the QHA,
obtained by taking into account the fluctuations around the so-called
flux phase \cite{Affleck,Kim}.
The situation differs considerably from the case of bosonic
matter fields because we have only one coupling constant. However,
the $\beta$-function of the $\alpha$-coupling has the same form as
in Eq. (\ref{betaalpha}), except for a different expression
for $\gamma_A$, which is here  a function of $\alpha$ alone.
All the preceeding results for the bosonic theory apply,
but there is no critical line and the fixed point $\alpha_*$ is
a function of $N$ only. Hence, we expect a critical 
value $N=N_c$ at which a CD transition takes place.
By a one-loop renormalization group calculation, 
we obtain 
$\gamma_a= N \alpha/8$, giving therefore the 
approximate value 
$ \alpha _*=8/N$.
Inserting this fixed point value into the 
sine-Gordon-like Lagrangian
and using the fact that $t_c=12\pi^2$,
we find a critical value
$N=N_c \equiv 24$ separating the
confined from the deconfined regime. 
This agrees with the rather crude value obtained 
in Ref  \cite{IL} but 
we can in fact expect a true $N_c << 24$ .
For instance, Marston computed the effective action for the  
monopoles approximately, obtaining the much lower value 
$N_c=0.9$ \cite{Marston}.  This shows that presently, there 
is considerable uncertainty in determining $N_c$. A precise determination
of $N_c$ is not the topic of this paper, but we shall now point out 
to the reader some subtleties concerning the fermionic case.  

In principle, the above results 
indicate that in the presence of  massless Dirac fermions, a
compact U(1) gauge field  would confine electric charges
for $N<N_c$ and  deconfine them for $N>N_c$. 
However, the above considerations are valid only 
in the absence of spontaneous
chiral symmetry breaking. If such symmetry breaking occurs, 
the fermions become massive
and no anomalous dimension is generated for the gauge field.
Chiral symmetry breaking is believed to occur for $N<{N}_{\rm ch}$,
where typically ${N}_{\rm ch}\in(3,4)$  \cite{Appel}. The dynamical
mass generation in Eq. (\ref{Dirac}) is usually shown
by using a Schwinger-Dyson approach controlled by a $1/N$
expansion \cite{Pisarski} and, therefore, is inherently non-perturbative.
If the true critical value of $N$ is such that  
$N_c<N_{\rm ch}$, then the value of $N_c$  should  be considered 
as a calculational artifact. This is because in our picture, 
this value  is a direct consequence of the existence of 
an anomalous scaling behavior for the gauge field which, as 
discussed above,  
does not exist if a fermion mass is spontaneously generated. 
A closely related argument is that the screening properties 
of the theory weaken  the logarithmic interaction, thus leading 
to a $1/R$ behavior of the interaction between monopoles 
\cite{Marston}. If such a  scenario were to hold, the monopoles would 
never be confined, and as a consequence no CD-transition would take 
place in the case of three-dimensional QED defined by Eq. (\ref{Dirac}).    

Summarizing, we have studied the influence of the gauge field
anomalous dimension induced by the coupling to matter fields to
the confinement-deconfinement transition. Our analysis reveals
that the anomalous scaling of the gauge field plays an essential
role for bosonic matter which possesses two relevant couplings.
There, the electric charges deconfine for
a Ginzburg parameter $\kappa>\kappa_c=1.17/\sqrt{2}$, inside the
type II regime of a superconductor. For the fermion theory, on 
the other hand, a deconfinement transition seems to take place only as a
function of the number $N$ of fermion components. However, due 
to the possibility of spontaneous chiral symmetry breaking and/or 
strong screening effects, further study is necessary in order 
to firmly establish that a deconfinement transition really takes 
place in the fermion theory. Such a study will be in part numerical, 
with the use of Monte Carlo simulations. Detailed large-scale
Monte Carlo simulations for the bosonic case are currently 
in progress \cite{Smiseth}.  

We thank J. B. Marston for stimulating discussions and
comments. We also thank C. Mudry for his interesting remarks. 
AS thanks the hospitality of the Institute of Theoretical Physics of the
Free University Berlin where this work has been done, and support from 
Norwegian Research Council through Grant 148825/432.
FSN  gratefully acknowledges  financial support 
from the Alexander von Humboldt
foundation.

\end{document}